\documentclass[aps,pre,twocolumn,groupedaddress,amssymb]{revtex4}
\usepackage{graphicx}
\begin{document}

\def\mdr{\mathit{\delta r}}
\newcommand{\aver}[1]{\langle #1 \rangle}

\title{Finite size effect of harmonic measure estimation in a DLA model: variable size of
probe particles}

\author{Anton Yu. Menshutin$^{1,a}$, Lev N. Shchur$^{1,2}$, and Valery M. Vinokour$^2$}

\affiliation{ $^{1)}$ Landau Institute for Theoretical Physics,
142432
Chernogolovka, Russia \\
$^{2)}$ Materials Science Division, Argonne National Laboratory,
Argonne, Illinois 60439, USA \\
$^a) $e-mail: \tt may@itp.ac.ru}

\begin{abstract}

A finite size effect in the probing of the harmonic measure in
simulation of diffusion-limited aggregation (DLA) growth is
investigated. We introduce variable size of the probe particles to
estimate harmonic measure and extract fractal dimension of DLA
clusters taking two limits, of vanishingly small probe particle size
and of infinitely large size of a DLA cluster. We generate 1000 DLA
clusters consisting of 50 million particles each using off-lattice
killing-free algorithm developed in the early work. The introduced
method leads to an unprecedented accuracy in the estimation of the
fractal dimension. We discuss the variation of the probability
distribution function with the size of probing particles.
\end{abstract}


\maketitle
\section{Introduction}

There is a wealth of processes and systems ranging from the crystal
growth and dendrites formation to bacteria colonies growth and
dielectric discharge patterns (see~\cite{BH} for a review) that
evolve in time in a similar manner following a generic process
called the \textit{two dimensional aggregate growth}.  The specific
model which is a most common representation of this process is the
diffusion limited aggregation (DLA)~\cite{WS} and its
generalization, dielectric breakdown model (DBM)~\cite{Pietronero}
capturing most of the dynamic properties of the random aggregates
evolution~\cite{BH}. While analytical and numerical studies
have had impressively advanced  our understanding of the
DLA~\cite{BH,TH-review}, several critical issues remain unresolved. One of the
major controversies is related to the multiscale and fractal nature
of DLA.  It has been a common belief that DLA, DBM and Laplacian
growth~\cite{Saffman58} belong to the same universality class.
Recently, however, this common wisdom was questioned in
Ref.~\cite{Barra2001,Hentschel2002}, where it was argued that DLA
and Laplacian growths have different fractal dimensions.  To resolve
this intriguing issue a breakthrough in the precision of the data on
fractal dimensions of both models is required.  In our work we will
address the algorithms and techniques that offer a dramatic
improvement in determining the DLA fractal dimension $D$
characterizing DLA clusters.

The DLA/DBM models and growth algorithms are based on the
aggregation of the particles randomly diffusing in the plane.  A
noticeable progress was made owing to the technique introduced
recently by Hastings and Levitov~\cite{HL} where the electric field
equations appearing in DBM were solved by iterative conformal
mapping, and freely diffusing particles could reach any given site
of the surface.  The approach of~\cite{HL} was further extended by
Hastings~\cite{Hastings02} where  the models of the Laplacian Random
Walk (LRW) similar to DLA, with the exception that the growth occurs
only at the tip, were proposed. In the continuum limit this model
transforms into the stochastic Loewner evolution
(SLE)~\cite{SLE,Cardy-SLE-rev}. Yet, within the conformal mapping
techniques fairly large clusters are to be built to insure a
reasonable precision in determining the fractal dimensionality
because the relative fluctuations decrease very slow with the
cluster size. Namely, squared relative fluctuations
\begin{equation}{\cal
F}_D{=}\left(\aver{D^2}{-}\aver{D}^2\right){/}\aver{D}^2
\label{D-var}
\end{equation}
\noindent  scale as $N^{-0.33}$ with the cluster size
$N$~\cite{RS,DLA-MS}. This is caused by large deviations in the
cluster structure across the ensemble: each cluster has its own
particular shape and preferred directions; these shapes vary in time
and do not converge to any common ``typical'' structure.

Random aggregate growth is fully characterized by the harmonic
measure, i.e. by the probability for the surface to advance at some
given segment.  A common recipe for measuring this probability in
computer experiments was to use the probe particles of the same size
as particles that comprise the aggregate.  To count the
controversies concerning the issue of the universality of random
aggregate growth, we undertake more precise analysis of DLA model,
based on the probing particles of variable size.  As it was
shown~\cite{DLA-MSV}, this technique can significantly increase the
accuracy of the fractal dimension estimation.  We build the ensemble
of 1000 DLA clusters, 50 millions particles each, generated by the
off-lattice killing-free algorithm~\cite{DLA-MS} which we now modify
to insure  the free zone tracking. By using probe particles of
different size $\delta$ and then taking the limit $D(\delta,N),
\delta \rightarrow 0$ we find fractal dimension measured on the
whole surface including also the regions that were inaccessible to
the probe particles of the fixed size.  Another important feature of
our approach, is that the small particles are more sensitive to the
fjords, while the large particles feel better the tips of the
cluster. Thus, using the variable size-particles we measure not only
the influence of the active growing zones (the hot part of the
cluster) but also the effect of the ``accomplished" (frozen, or
cold) parts of the cluster. As a result, the large fluctuations of
the fractal dimension $D(\delta,N)$ are suppressed. Taking then the
second limit $D=D(N), N \rightarrow \infty$ we find~\cite{DLA-MSV}
that $D$ converges to the value of $D=1.7100(2)$ which is the order
of magnitude improvement as compared to past simulations.

The paper is organized as follows.  In section~\ref{sec-simulations}
we give details of the off-lattice killing-free algorithm we use to
generate clusters and of the procedure of estimation of DLA fractal
dimension. Section~\ref{sec-oscill} contains brief discussion of the
dependence of fractal dimension from the cluster size and
section~\ref{sec-cm} discusses walk of the center of cluster mass as
cluster grows, and the two ways to calculate cluster radii - as
distance from origin and as distance from center of mass.
Section~\ref{sec-probe} describes the original method of fractal
dimension estimation with the additional parameter, the size of the
probe particles. We discuss in section~\ref{sec-pdf} how the
probability distribution function for particle to stick cluster at
the distance $r$ varies with the size of the probe particles.
Finally, in the section~\ref{sec-disc} we discuss our results and
possible future developments.

\section{Simulations}
\label{sec-simulations}

\subsection{DLA algorithm}

The original DLA algorithm~\cite{WS} starts with placing a seed
particle at the origin of a square lattice. Than at certain
position, far away from the seed, a new particle is released, which
wonders stochastically over the lattice until it runs into and
sticks to the seed particle. If in the course of random walk a
particle crosses the lattice boundary it gets removed (killed). A
new particle is generated and process repeats. Successive
aggregation of particles forms a DLA cluster. Several improvements
to this original algorithm were developed in order to speed up the
simulations ~\cite{BB-alg,TM-alg}. Later it was found that large
cluster exhibit anisotropy which reflects the symmetry of the
lattice~\cite{MBRS-alg}.

There exist also the off-lattice versions of the same algorithm.
Namely, the particles are assumed to be balls that move freely in
each direction. Making use the off-lattice algorithm one can easily
speed up simulations by utilizing improvements introduced
in~\cite{BB-alg,TM-alg,KVMW,SSZ,book}, in particular, the large step
sizes, returns instead of killing, etc.

Another algorithm for generating the DLA-like cluster is the
well-known DBM model~\cite{Pietronero}. Instead of tracing the
motion of particles, one solves a  Laplace equation and calculates a
probability for a particle to hit the surface at the given site.
Adding particles successively  with the calculated probability, one
generates a DBM cluster.

\subsection{Off-lattice killing-free algorithm}

We utilize the realization of the off-lattice killing-free algorithm
described in~\cite{DLA-MS}, in which we upgraded the free-zone
calculation. An essential feature of our algorithm, that it is a
memory saving one; the memory organization is similar to that used
in the Ball and Brady algorithm~\cite{BB-alg}.  The modifications we
have introduced are as follows: (i) we employ the large walk steps;
(ii) we use a killing-free rule for exact evaluating the probability
for the particle to return at the birth circle; (iii) we use only
two layers in the memory hierarchy; (iv) we use a recursive
algorithm for the free zone tracking; and (v) the size of free zone
is calculated precisely for the particles moving near the cluster
boundary. Our further improvement of the walk procedure [item (iv)
above] is that instead of performing the search of free cells each
time particle is moving, we pass all the cells and recalculate the
free-zone sizes after the addition of each particle.

\begin{figure}
\includegraphics[angle=0,width=\columnwidth]{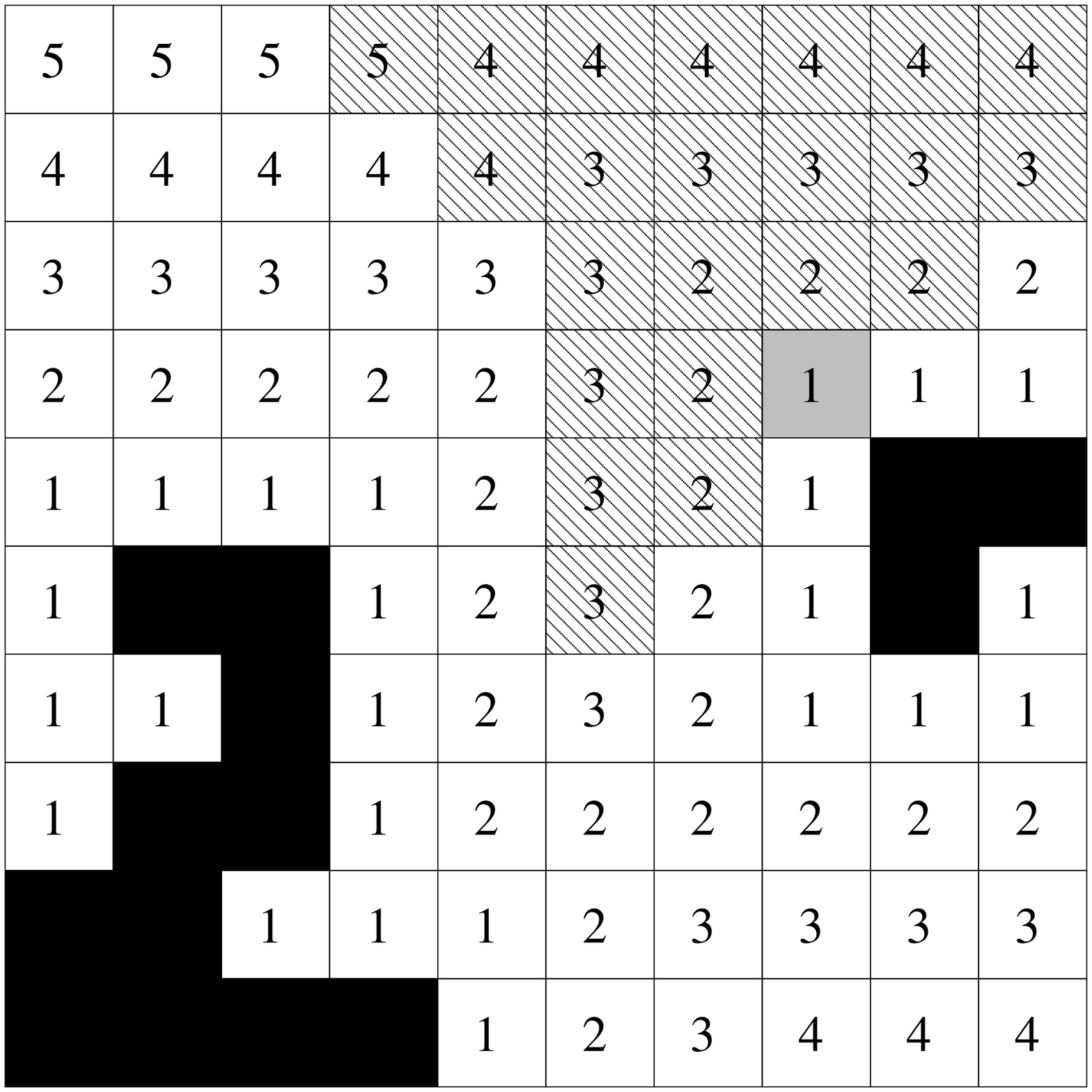}
\caption{Occupied lattice sites and sizes of free zones around
each cell. The gray cell is the newly occupied one. Shaded
cells are cells where free zone size will be recalculated.
}
\label{lattice-pic}
\end{figure}

The killing-free algorithm consists of the following steps: 1)
Placing a seed particle at the origin (0,0). 2) Generating a new
particle at a random position on $R_b$. 3) Moving the new particle
by fixed steps in random directions. 4) Aggregating a particle to
the cluster in case it collides with either the seed or the cluster
itself. 5) Generating a new particle position on $R_b$ with the
probability
$$P(\delta\phi)=\frac{1}{2\pi}\frac{1-x^2}{x^2-2x\cos{\delta\phi}+1},$$
where $x=r/R_b$ and $\delta\phi$ is the change in particle angle, in
case where the particle falls into a dead circle $R_d$, $R_d>R_b$.
6) Repeating the cycle from the step 2.

Since particles are moving randomly in the plane and assuming that
the step size is negligible as compared to the dimension of the free
zones around a particle (unless it gets close to the cluster), one
could speed up the computational process by increasing the step
size. This requires reliable and efficient identification of such
``cluster free" regions. The common approach is to use the
hierarchical memory model and store particles into cells, which will
comprise the first layer. The second layer consists of cells which
are several times larger than those at the first layer, and so on.
Starting with the largest-scale cells layer, one checks their
occupancy until the unoccupied cell is found. If the process is
successful one gets the size of free zone at ones. Unfortunately,
for the large enough clusters this approach is memory consuming and
inefficient, because of the high degree of the granularity.

We use a double layer memory organization: Particles are stored in
the square cells, the latter also describe the size of free area
around them.  This information is updated during the attaching a new
particle. Let $i,j$ be the indexes of some cell on the lattice and
$x_n,y_n$ be the indexes of some occupied cell. Then the free zone
around the $i,j$ cell is $d_{i,j}=\min_n{max(|i-x_n|,|j-y_n|)}$
where the $\min$ is taken over all occupied cells.
Figure~\ref{lattice-pic} shows the lattice where occupied cells are
black and numbers denote the free area sizes around each empty cell.
As a new cell gets occupied (marked grey in the Figure) one has to
check all the cells around it and recalculate $d_{i,j}$. Shaded
cells on the picture are those where the distances are going to be
changed. One can see that the perturbation propagates continuously
-- all the shaded cells are connected with each other. This
continuous propagation holds for every configuration of cells.

In order to track down all the cells where the distances have
changed, one should traverse first all the adjacent cells around the
new one (grey cell in the Figure), then the next nearest cells that
have changed the neighbors and so on.  Since this process is
unrestricted in time and can continue indefinitely (corresponding to
the upward propagation in the figure), we stop the process after
having spanned the distance of 15 and assume that cells where
distance is not specified have free zone sizes of 15 as well.

For large clusters the cavities between the branches become bigger
than 15 cell sizes. In order to overcome this limitation we use the
two layer memory model each having free zone size counters updated
by the rules we have described above. For clusters of size 50
million particles we choose first layer cell size of 32 and second
layer - 128, and the particle radius is unity.

On the close distance to the cluster we found useful to find precise
distance to the cluster by iterating over all adjacent particles.
This step is only required when current cell is marked as occupied
or is adjacent to the occupied one and big step based on coarse
grain information given by free/occupied cells is not enough.

All these improvements of DLA realization algorithm enable us to
generate each cluster with 50 million particles in about 3 hours on
3Ghz Pentium 4 with 2 GB of RAM.

\subsection{Fractal dimension estimation}
The fractal dimension of a random aggregate is the quantity $D$
defined by the scaling relation

\begin{equation}
R \propto N^{1/D}, \label{ens-D}
\end{equation}

\noindent where $R$ is a linear measure of the cluster consisting of
$N$ particles. Several possible choices of the characteristic length
$R$ are listed in Table~\ref{tab-fractal-dimension}.

\subsubsection{Ensemble averaging}
Let $M$ denote the number of clusters we have generated and $r_N(k)$
be the position of the $N$-th particle in the $k$-th aggregate. Then
the deposition radius is
$$R_{\rm dep}(N)=\aver{r_N}= \frac{1}{M} \sum_{k=1}^{M} r_N(k).$$
Other quantities we are interested in are the gyration radius,
$R_{\rm gyr}$, the root mean squared (RMS) radius $R_2$, and the
penetration depth $\xi$, with the definitions given in the upper
part of the Table~\ref{tab-fractal-dimension}.

In order to understand the behavior of the fractal dimension in the
limit of infinite clusters one can check the dependence of $D$
versus $N$. We extract $D(N_0)$ as the result of fitting some $R$ to
the scaling relation~(\ref{ens-D}) when $N$ varies from 0 to $N_0$
or in other words using clusters of size less than $N_0$.

\subsubsection{Harmonic measure averaging}

Harmonic measure is defined on the surface of the cluster as the
probability of the DLA cluster to grow at the given point. One can
define the deposition radius  as $R_{\rm dep}^{hm}=\int rdq$, where
the integral is taken over the cluster surface and $dq$ is the
growth probability at point $r$. Using this harmonic measure we
calculate fractal dimension of $k$-th cluster $D_k(N)$, and perform
averaging over the ensemble of clusters,
$D(N)=\frac{1}{M}\sum_{k=1}^M D_k(N)$.

In simulations we estimate harmonic measure averages in the
following way. We use probe particles which move due to the
convention rules but do not stick to the cluster. Instead we track
the position $r_i$ where $i$-th particle hits the cluster. Then
deposition radius $R_{\rm dep}^{hm}$ estimated as average $R_{\rm
dep}^{hm}=\frac{1}{N_p}\sum_{i=1}^{i=N_p}r_i$. Calculating
harmonic-measure averages is very time consuming procedure. The
number of probe particles $N_p$ used is chosen dynamically so that
relative error of $R_{\rm dep}^{hm}$ is less than 0.001. See the
lower part of Table~\ref{tab-fractal-dimension} for the definitions
used.

\begin{table} \begin{tabular}{|l|l|l|l|}
\hline
\multicolumn{4}{|c|}{Ensemble averages} \\
\hline
&Length & Definition & $D$  \\
\hline
Deposition radius & $R_{\rm dep}$  & $\aver{r} $              & $1.71111(59)$ \\
RMS radius        & $R_2$          & $\sqrt{\aver{r^2}}$      & $1.71155(29)$ \\
Gyration radius   & $R_{\rm gyr}$  & $\sqrt{\frac1N \sum_{k=1}^N
                                    \aver{r^2}_k}$            & $1.71149(30)$ \\
Penetration depth & $\xi$          & $\sqrt{{R_2}^2-{R_{\rm
dep}}^2}$
                                                            & $1.7184(65)$  \\
\hline
\multicolumn{4}{|c|}{Harmonic measure averages} \\
\hline
Deposition radius &$R_{\rm dep}^{hm}$& $ \int r\, {d}q$& $1.70922(97)$ \\
RMS radius        &$R_2^{hm}$          & $\sqrt{ \int r^2\, {d}q}$
                                                            & $1.70944(87)$ \\
Effective radius  &$R_{\rm eff}^{hm}$  & $\exp{(\int \ln r\, {d}q)}$
                                                            & $1.70944(87)$ \\
Penetration depth &$\xi^{hm}$          & ${R_2^{hm}}^2-{R_{\rm
dep}^{hm}}^2$
                                                            & $1.74(3)$     \\
\end{tabular}
\caption{Estimates of the fractal dimension $D$ extracted with the
fit $N \propto R^{D}$ to the dependence of the various lengths with
ensemble size of 1000 clusters each with $5\cdot10^7$ particles.}
\label{tab-fractal-dimension}
\end{table}

\section{Oscillations of fractal dimension with the cluster size and weak self-averaging}
\label{sec-oscill}

One can expect that the fractal dimension can be found as the limit
$D(N)$, as $N \rightarrow \infty$. Unfortunately, this simple
approach fails because of the non-monotonic $D(N)$ dependence with
the rather large variations in $D(N)$~\cite{DLA-MS}.

\begin{figure}
\includegraphics[angle=0,width=\columnwidth]{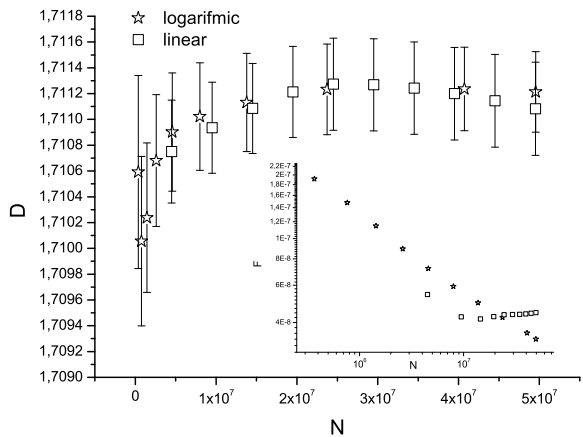}
\caption{Fractal dimension as function of $N$ measured with $R_{\rm
dep}^{hm}$ and harmonic measure averaging over 100 clusters of size
$5\cdot10^7$. Stars represent estimation in logarithmic scale, and
boxes - in linear scale. In the inset: dependence of relative
squared fluctuations of $D$ with $N$.}
 \label{d_n_fluctuation}
\end{figure}

We measure fractal dimension for each cluster using $R_{\rm
dep}^{hm}$ and then we calculate the ensemble average of fractal
dimension and its variation. There are two ways of extracting $D$
from the $R(N)$ dependence: either via choosing points in $R_{\rm
dep}^{hm}$ curve to be uniform on either the linear scale, or on the
logarithmic one. It seems that both procedures result in the same
average value of $D$, but exhibit different behaviors of the
variances, see Figure~\ref{d_n_fluctuation}. If points are
distributed uniformly in the logarithmic scale, then $\cal
F_D$~(\ref{D-var}) changes as $N^{-0.33}$ as was found previously in
our paper~\cite{DLA-MS}. In contrast, $D$ measured by the points
uniformly distributed on the linear scale shows larger fluctuations
without any visible decay with $N$.

One can try to understand this behavior as follows. Varying points
distribution we, accordingly, change their respective weights. If
points are uniform on the linear scale, then $D$ is more sensitive
to the behavior of the bigger clusters, and the error when
determining $D$ does not change with the size of the cluster very
much. One can say that $D$ exhibits the lack of self-averaging in
this case. On the contrary, the uniformity on the log scale gives
rise to a weak self-averaging of $D$.

The two procedures both yield very close values of $D$ as one can
check in Figure~\ref{d_n_fluctuation}, therefore for the rest of
this paper we will use the points uniformity on the linear scale
because this choice requires less CPU time for calculating the
harmonic measure.

\section{Center of mass fluctuations}
\label{sec-cm}

\begin{figure}
\includegraphics[angle=0,width=\columnwidth]{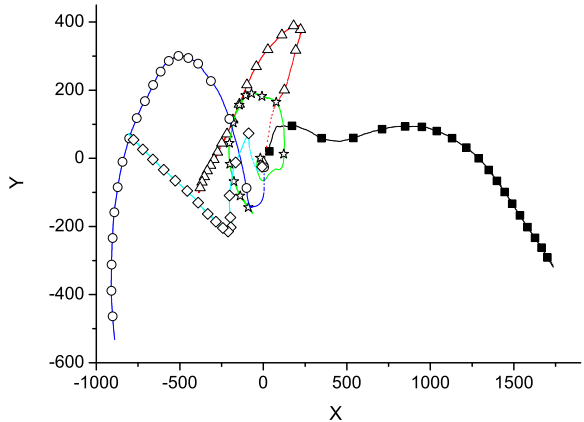}
\caption{(Color online) Center of mass position of five different clusters growing
from $10^5$ to $5\cdot10^7$ particles, with the interval of $2\cdot10^6$. 
See discussion for details.}
\label{cm-position}
\end{figure}

Each realization of a DLA cluster is a unique object: its branch
structure varies very much from sample to sample.  This causes large
fluctuations in cluster properties and reflects in the behavior of
the center-of-mass position of a cluster. At any moment of time
(which is proportional to the number of particles in the cluster
$N$) there are some preferred directions of growth. These directions
vary in time as illustrated in Figure~\ref{cm-position}, where we
present the variations of the center-of-mass positions in plane
coordinates $(X,Y)$ for five different clusters. The center-of-mass
positions (CoM) for each cluster are marked with their own symbol
(circles, rhombi, triangles, stars, and boxes). Each mark placed
after $2\cdot10^6$ particles were added, so the history of the center-of-mass
variation represents the time from $10^5$ to $5\cdot10^7$. One sees from
Figure~\ref{cm-position}, that for some clusters (marked with
stars), position of the CoM is mainly rotated around the origin,
while for some of them (marked with boxes) CoM is diffused from the
origin in the given time interval.

On average, clusters are quite uniform when looking for the average
displacement of CoM. We estimate the average angle
$\aver{\phi}=-0.01504998(\pm 0.05535)$ of CoM (in the interval
$[-\pi;\pi]$) to be close to zero, while the angle variance is
approximately $\sqrt{\aver{\phi^2}-{\aver{\phi}^2}}=1.76$. This is
very close to the value one can find assuming that $\phi$ is
uniformly distributed in given interval: $\aver{\phi}=0$ and
$\sqrt{\aver{\phi^2}-{\aver{\phi}^2}}=\pi/\sqrt{3}\approx 1.81$.

\begin{figure}
\includegraphics[angle=0,width=\columnwidth]{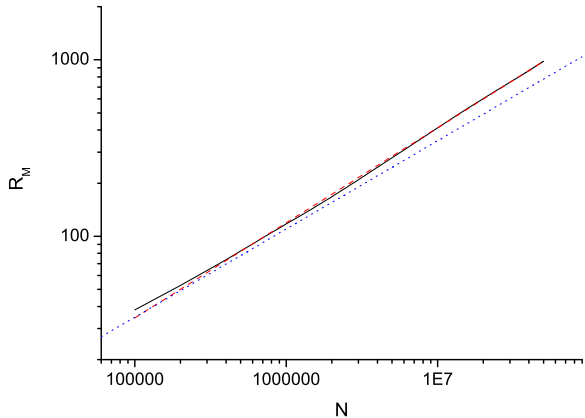}
\caption{(Color online) Dependence of center of mass position
$R_M(N)$ with the cluster size $N$, as calculated averaging over
1000 cluster with $5\cdot10^7$ particles. Dashed line is a fit to
$R_M(N)$ and dotted line have a slope 1/2. } \label{rm-position}
\end{figure}

There have been several attempts in the literature to relate the CoM
distance from the origin $R_M$ to the expression~(\ref{ens-D}),
defined as
$$R_M(n)=\sqrt{\frac{1}{K}\sum_{k=1}^K\left(\frac{1}{N}\sum_{i=1}^N(\vec{r_i})\right)^2},$$
\noindent where $K$ is the number of clusters. We, however, find no
reason for such a connection, because $(X,Y)$ changes rather
randomly. One can think of it as of a random walk in the plane,
reflecting the competition between the growing activity of the
branches.  One thus expects the corresponding exponent to be equal
to $d=2$ (and not to $D\approx 1.71$) in the limit of the infinite cluster size.
Yet, the rare events such as the active branch growth effectively
lower this value. We have found, that at large cluster sizes the
generated exponent may be estimated as $1.8558(13)$ which is notably
larger that $D\approx 1.71$.  The  dependence of $R_M(N)$ is plotted
in Figure~\ref{rm-position} with solid black line, the dash line is
the fit, and dots line corresponds to the slope equal to 1/2
 (reverse exponent).

One can calculate the radius of the cluster   by either (i) choosing
the origin (0,0) as a reference point, or (ii) by choosing a
temporary CoM position in the $(X,Y)$ plane  as a reference point in
the estimation of radii given in Table~\ref{tab-fractal-dimension}.
It turns out that the fractal dimension calculated by means of
either of the recipes is the same within the computational error as
one can see comparing the second and third columns in the
table~\ref{tab1}. However, this effect shows the additional possible
reason for high fluctuation rates in DLA cased by the unique random
branch structure of each cluster.

\begin{table}
\begin{tabular}{|l|l|l|} \hline
 & $r$ to seed & $r$ to center-of-mass   \\ \hline
 $R_{\rm dep}$ & 1.71111(59) & 1.71112(60) \\
 $R_2$   &   1.71155(29) & 1.71149(54) \\
 $R_g$   &   1.71149(30) & 1.71133(30) \\ \hline
 \end{tabular}
\caption{Fractal dimension estimated by the standard method using
the stick distance $r$ to the seed (second column) and to the
center-of-mass (third column) as estimated for the number of typical
lengthes.} \label{tab1}
\end{table}

\section{Fractal dimension estimation: variable probe particles}
\label{sec-probe}

A common technique of enhancing the precision of computations
is the noise reduction in DLA. In doing so, one selects only the
most probable events by attaching the hit counters to each particle
of the cluster. The new particle is added to the particular old
particle when the old particle is hit with some prescribed number of
times.  Similarly, the same procedure can be used when calculating
the harmonic measure averages. Tips are the most probable places to
grow, and at the same time the fluctuations at the tips are more
strongly associated with the drastic changes in the cluster
geometry, which is reflected in the competing growing of the
branches, in the birth of the new branches, and in the mutual
screening of branches. The internal (frozen) part of the cluster
surface, which is screened in the fjords and carry the small part of
the measure, may give some contribution to the harmonic averages due
to its length which can be large. So, we suppose that the
internal part of the DLA cluster may contribute to the average. To
check on this point, we introduce the variable size of the probe
particles. We choose the size of the particles which we used to
build the cluster to be the unit of the length, and measure in this
units the probe particles size, $\delta$. Larger particles increase
the growth probabilities at the tips, while smaller particles will
penetrate dipper in the fjords, and redistribute harmonic measure
shifting the part of the growth weight from the tips to the fjords
(see next Section for details).

\begin{figure*}[htb]
\includegraphics[width=\textwidth]{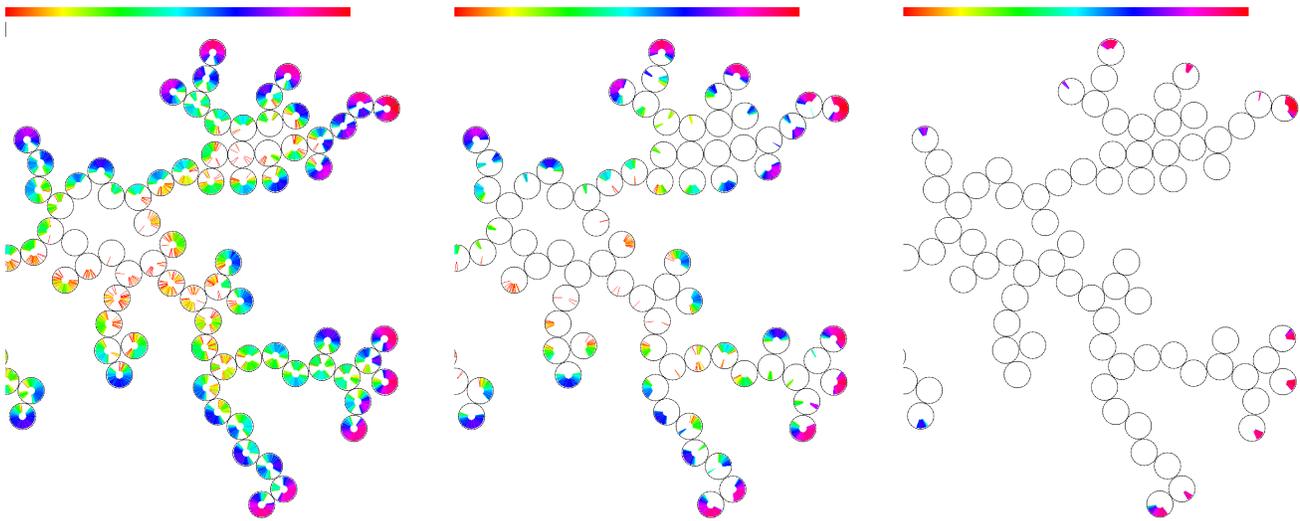}
\caption{(Color online) Fragment of the cluster with $N=10^3$
particles. Intensity of the segments is proportional to the
logarithm of the hit probabilities, corresponding to the probe
particles of size $\delta=0.1,1$, and $10$, from left to right.
Probabilities are larger near the tips and smaller inside the
fjords. For lower $\delta$ the bigger part of the cluster surface is
touched by probe particles. } \label{pic-surface}
\end{figure*}

Figure~\ref{pic-surface}  illustrates the variation of the  harmonic
measure estimation with the probe particle size $\delta$. Here the
intensity reflects the probability (in the log scale) for the
particle to hit the given segment of the cluster. The probe
particles of the size $\delta=10$ hit only few particles at the tips
and never go inside the fjords. Only this hottest part of the
cluster contributes to the harmonic measure estimation. The
particles of the size $\delta=1$ touch the bigger part of the
surface but still there are regions inaccessible to them. The
smaller particles of the size $\delta=0.1$ (recall that this size is
less than that of the building particles of the size 1) touch
visibly the larger part of the cluster surface, penetrating deeper
into the fjords.

Two features contribute to the fjord screening: the size of a probe
particle and the bottleneck configuration. Even if a particle fits
into the bottleneck, the probability for it to pass is very small if
the size of the particle is comparable to the bottleneck width. In
other words, the effective channel width is somewhat narrower than
the bottleneck width. Using the variable size probe particles we
suppress the influence of the bottleneck and can thus measure the
lower probabilities at the fjord surface. At the first glance, our
approach looks opposite to the noise reduction method, which cuts
the parts of the surface with the low harmonic measure. In fact,
this contrast is misleading, because noise reduction method affects
the {\it growth process} of the cluster, and our method affects the
{\it measurement process} of the harmonic measure, and leads to the
higher precision of the estimation of harmonic averages.

One can think of our approach as of the canonic measurement of the
length of the coast: the smaller the ruler, the bigger the coast
length is~\cite{Mandelbook}. In our case,  the smaller probe
particles, the longer the reachable surface is, with the cut-off on
the scale of the bulk particles.

\subsection{Number of particles on the surface}

\begin{figure}
\includegraphics[angle=0,width=\columnwidth]{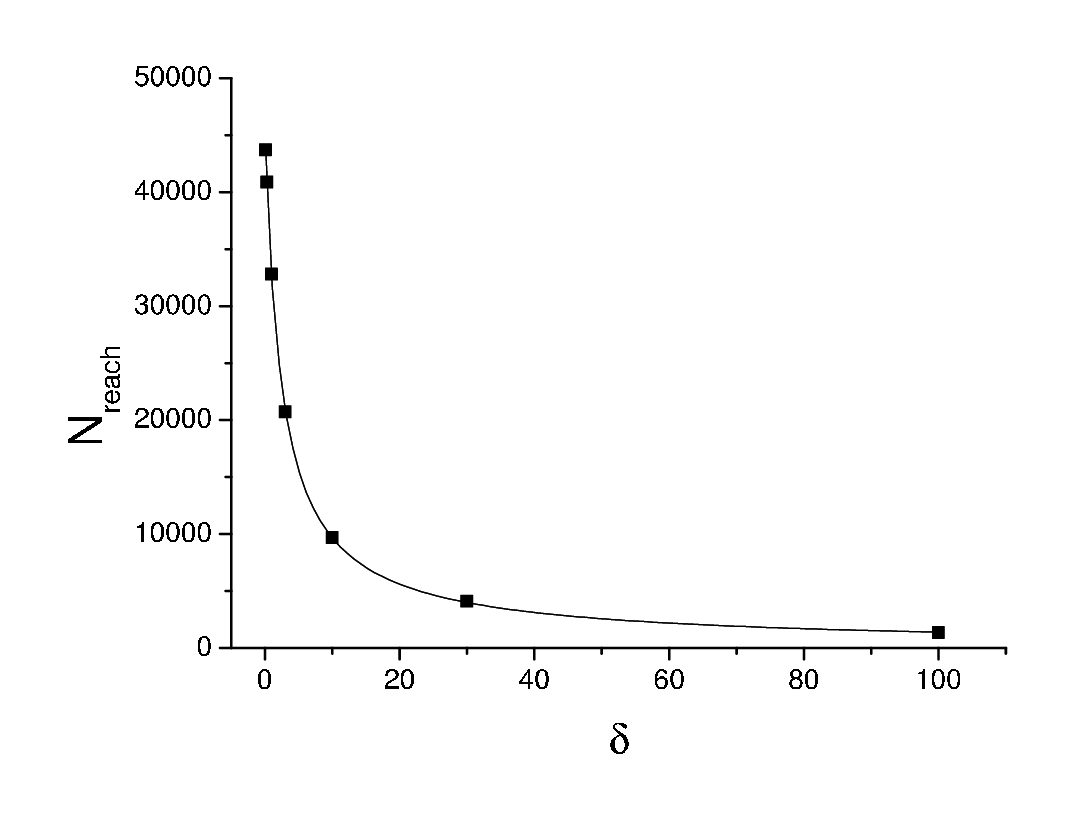}
\caption{Number of particles on the surface reachable by probe
particles of size $\delta$. Measured by $10^6$ probe particles and
clusters of size $10^6$.} \label{pic-surface-n}
\end{figure}

By varying the size of probe particles we change the reachable
effective cluster surface. The number of particles on the surface
reachable during the probing process is shown in
Figure~\ref{pic-surface-n}. Solid line on the picture is the fit to
the expression

\begin{equation}
N_{reach}=N_{surf}/(1+\delta/\delta_0)^\alpha \label{N-reach}
\end{equation}

\noindent with $\alpha=0.91(1)$, $\delta_0=2.23(6)$ and
$N_{surf}=45697(187)$. $N_{surf}$ is a total number of surface
particles in cluster, those reachable by probe particles with
infinitesimal size, and clusters size fixed to $10^6$. Thus, in the
measurements with $10^6$ probe particles one can touch only about
$N_{surf}\approx 45700$. The number of reachable particles for
$\delta=1$ is lower by the factor of $1.4$ than for particles of
size $\delta=0.1$. The number of reachable particles is the complex
function of the probe particle size and of the number of probe
particles. The best way to fill the whole surface of the cluster
with probes is to find the limit $N_{probes} \rightarrow \infty$.
But this is very time consuming procedure.

\subsection{$D$ in the limit of vanishing probe particle size}

The fractal dimension of the aggregate turns out to depend also on
$\delta$, and the effective fractal dimension is now a function of
two variables, $D(N,\delta)$. We are interested in the double limit
$D=\lim_{N \rightarrow \infty,\delta \rightarrow 0}D(N,\delta)$. Let
us first find the limit of the vanishing $\delta$, which gives us
the $D(N)$ dependence:

\begin{equation}
D(N)=\lim_{\delta\rightarrow 0} D(N,\delta) \label{D_N}
\end{equation}

\noindent with the help of the fit

\begin{equation}
D(\delta,N)=D(N)+A\delta^\beta. \label{DdN}
\end{equation}

\begin{figure}
\includegraphics[angle=0,width=\columnwidth]{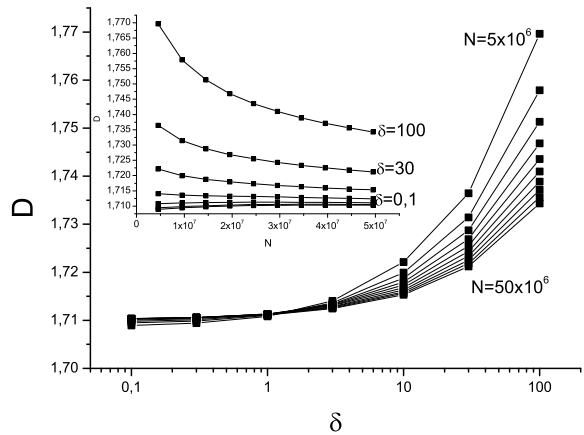}
\caption{Fractal dimension $D(\delta,N)$ as a function of the probe
particle size $\delta$ for different cluster size. Inset:
$D(\delta,N)$ is shown as a function of $N$ for the set of  $\delta$
values. Lines are the guide for the eye. See text for details.} \label{d_delta}
\end{figure}

We calculated harmonic measure averages with the set of probe
particles sizes: $\delta=0.1,$ 0.3, 1, 3, 10, 30, and 100. The
resulting fractal dimension $D(\delta,N)$ is shown in
Figure~\ref{d_delta} as function of $\delta$ for $N=5$, 10, 15, ...,
50 millions of particles. Fractal dimension $D(\delta,N)$ as a
function of N  for the set of $\delta$ is shown in the inset of
Figure~\ref{d_delta}, the topmost curve corresponds to $\delta=100$,
 while the lowermost - to $\delta=0.1$.

\begin{figure}
\includegraphics[angle=0,width=\columnwidth]{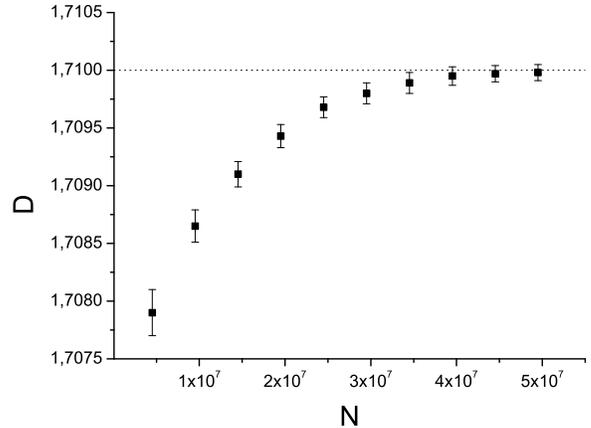}
\caption{Fractal dimension as function of $N$ for the limiting value
of $\delta=0$. It reach value $D=1.7100(2)$ (plotted with the dotted
line) in the limit of large cluster size $N$.} \label{dn-pic}
\end{figure}

Result of the data fit to Expr.~(\ref{DdN}) are shown in
Figure~\ref{dn-pic}.

Surprisingly, in the limit of the vanishing size of probe particles
fractal dimension become monotone function. As $N$ goes to infinity
$D(N)$ reaches its asymptotic value with approximate behavior (fit
using the last 6 data points) $D(N)=D+4.2\cdot 10^{12}/N^{2.17}$ which
gives the high precision of fractal dimension estimation
$D=1.7100(2)$.

\subsection{Errors estimation}

Making use of the least square fitting $\log(R_{\rm dep}^{hm})$ to
$\log N$ when calculating the fractal dimension of a single cluster 
$D_k(N,\delta)$, one can estimate an error on its determining.
As has been mentioned above, the clusters are unique, 
and this results in large fluctuations of
the fractal dimension over ensemble. This means that the error 
in the $D(N,\delta)$ averaged over the ensemble depends
mostly on
its variation, while the errors in each $D_k(N,\delta)$ 
are several times smaller.
The next stage of out calculation 
is constructing the nonlinear square fitting (NLSF)
to the equation~(\ref{DdN}) 
where all the data points have a weight, iversely proportional to the
error, assosiated with the point (instrumental error in terms of Origin
program). Resulting $D(N)$ curve is then fitted by the expression
$D(N)=D+A/N^B$ and the error for $D$ is again estimated with the help of NLSF.

\section{Probability distribution function}
\label{sec-pdf}

Harmonic-measure averages can be written in a form $R_{\rm dep}=\int
P(r,N)rdr$, where $P(r,N)$ is the probability for a particle to
stick to the cluster with $N$ particles at the distance $r$ from the
origin (or from center-of-mass as discussed in
Section~\ref{sec-cm}). This probability acquires a Gaussian
form~\cite{ms-PZ,nms-SBBS,Lee} upon averaging over the ensemble of
clusters.  It is not so if one calculates $P(r,N)$ at the surface of
a single cluster realization; rare events associated with the
particular shape contribute essentially to the probability function.
It was proposed in~\cite{DLA-MS} that fluctuations in $P(r,N)$ are
responsible for multiscaling issue.

\begin{figure}
\includegraphics[angle=0,width=\columnwidth]{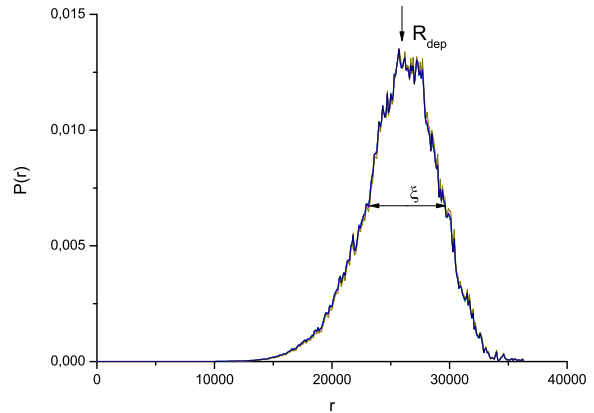}
\caption{Probability to stick at distance $r$ for different $\delta$
averaged over 1000 clusters with size $N=2\cdot 10^7$.}
\label{prob-pic}
\end{figure}

We measure $P(r,N,\delta)$ for a fixed cluster size $N=2\cdot 10^7$
and  the size of the probe particles $\delta$, and average over the
ensemble of 1000 clusters. The resulting function
$\overline{P(r,N)}$ is shown in Figure~\ref{prob-pic}. On the scale
of the figure, all lines practically coincide. Fluctuations of the
measured probability around the Gaussian form are visible despite
the large number of clusters we use for averaging.

The effect of the variation in the probe particle size on the
probability function may be investigated with the use of the
difference,

\begin{equation}
\delta P(r,N)=P(r,N,\delta)-P(r,N,\delta\rightarrow 0), \label{dPrN}
\end{equation}

\noindent which we approximate with the difference taken for the
smallest used probe particle size $\delta=0.1$, $\delta
P(r,N)=P(r,N,\delta)-P(r,N,\delta=0.1)$.

\begin{figure}
\includegraphics[angle=0,width=\columnwidth]{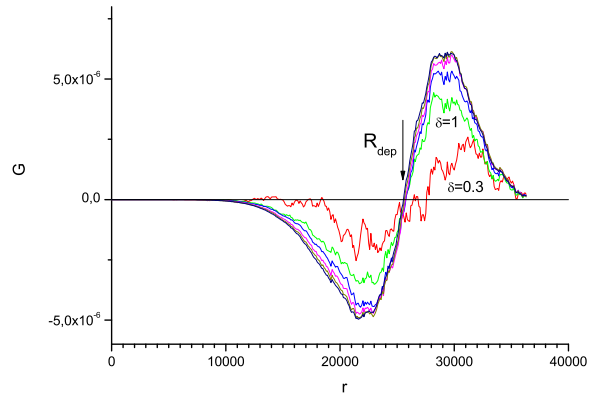}
\caption{(Color online). Normalized change of probability to stick at distance
$r$ for different $\delta$ relative $\delta=0.1$. Fixed point is at
$R_{\rm dep}$. Clusters size $N=2\cdot 10^7$.}
\label{delta-prob-pic}
\end{figure}

The variation $\delta P(r,N)$ demonstrates the sin-like shape, with
zero variation for any $\delta$ at the distance corresponding to the
average deposition radius $R_{\rm dep}$. The finiteness of the probe
particles leads to the increase of the probability for $r > R_{\rm dep}$
and probability decreases at distances $r < R_{\rm dep}$ (shown with
arrow in Figures~\ref{prob-pic} and \ref{delta-prob-pic}. The
amplitude of $\delta P(r,N)$ varies with $\delta$ as
$\delta^{0.78(1)}$. So we can normalize functions $\delta
P(r,N;\delta)$ with the multiplier $\delta^{-0.78}$,
$G(r,N,\delta)=\delta P(r,N,\delta) \delta^{-0.78}$. The set of this
functions is presented in Figure~\ref{delta-prob-pic}. All of them
completely coincide well. The only exception is for $\delta=0.3$ but
this is clearly due to approximation $P(r,N,\delta\rightarrow0)\approx
P(r,N,\delta=0.1)$ used in expr~(\ref{dPrN}). It seems, therefore,
that finite size of the probe particles will case the systematic
overestimation of the fractal dimension.

\section{Discussion}
\label{sec-disc}

Utilizing the variable probe particles for the harmonic measure
estimations allows analysis of fractal dimension as a function of
two variables. Fixing cluster size and taking the limit of the
vanishing size of a probe particle effectively smoothes out the
effects of the sample fluctuations as is seen from the monotonic
variation of the fractal dimension with the growing cluster size.
This contrasts favorably the conventional methods where the fractal
dimension can oscillate with the cluster size making the convergence
of the results very slow.  As a result using variable size
particles, one can extract the value of the fractal dimension from
$D(\delta,N)$ with the unprecedented high precision: we have found
that $D=1.7100(2)$ for the off-lattice DLA cluster in the plane.

We like to stress the accuracy of the above estimations as compared
to those obtained by the traditional methods mentioned in
Tables~\ref{tab-fractal-dimension}~and~\ref{tab1}. Technically, all
the estimations are compatible within the $2\sigma$ accuracy and, as
it seems, may be treated with the equal confidence. The major
difference of the values estimated with the help of varying probe
size, is that it is estimated not as a single random variable but as
the asymptotic value of some regular function as shown in
Figure~\ref{dn-pic}. Our approach corresponds to the variation of
the ruler scale in the conventional estimation of the fractal
dimension~\cite{Mandelbook}.

The major result of our work is that for the first time we have
reached a quality of the data which can offer reliable quantitative
answers. In particular, the obtained value of fractal dimension
allows to rule out the d=17/10 hypothesis proposed by Hastings about
10 years ago~\cite{Hastings97}. Furthermore, our technique can be used for the
analysis of fractal dimension as a function of the
lattice symmetry (d=3,4,5,6,7,8 and off-lattice corresponds to
infinity). The next interesting question to address is the
distribution of the harmonic measure near the tips and inside the
fjords, where one can expect our approach to be most effective.

\section{Acknowledgements}
This work was supported by the U.S. Department of Energy Office of
Science through contract No. DE-AC02-06CH11357 and the Program.
A.Yu.M. thanks Prof. G. Eilenberger and Landau Scholarship Committee
for support and Prof. H. M\"uller-Krumbhaar and Prof. E. Brener for
the kind hospitality and  useful discussions.


\begin{thebibliography}
\frenchspacing

\bibitem{BH} A. Bunde and S. Havlin, eds.,
  {\it Fractals and Disordered Systems}
(Springer, Berlin, 1996).

\bibitem{TH-review} T. C. Halsey, Physics Today {\bf 53} (2000) 36.

\bibitem{WS} T.A. Witten and L.M. Sander, Phys. Rev. Lett. {\bf 47}
(1981) 1400.

\bibitem{Pietronero} L. Niemeyer, L. Pietronero, H.J. Wiesmann,
Phys. Rev. Lett., {\bf 52} (1984) 1033.

\bibitem{Saffman58} P.G. Saffman and G.I. Taylor, Proc. R. Soc.
London A {\bf 245}, 312 (1958).

\bibitem{Barra2001} F. Barra, B. Davidovitch, A. Levermann, and I.
Procaccia, Phys. Rev. Lett. {\bf 87} 134501 (2001).

\bibitem{Hentschel2002} H.G.E. Hentschel, A. Levermann, and I. Procaccia
Phys. Rev. E {\bf} 66, 016308 (2002).


\bibitem{HL} M.B. Hastings and L.S. Levitov, Physica D {\bf 116} (1998) 244.

\bibitem{Hastings02} M.B. Hastings, Phys. Rev. Lett. {\bf 88} (2002) 055506.

\bibitem{SLE} O. Schramm, Israel J. Math. {\bf 118} (2000) 221; G.F Lawler, P. Schramm, and W. Werner,
Acta Math. {\bf 187(2)} (2001) 237; ibid. {\bf 187(2)} (2001) 275.

\bibitem{Cardy-SLE-rev} J. Cardy, Annals Phys. {\bf 318} (2005) 81.

\bibitem{RS} T.A. Rostunov and L.N. Shchur, JETP 95, 145 (2002).

\bibitem{DLA-MS} A. Yu. Menshutin and L.N. Shchur, Phys. Rev. E {\bf 73} (2006) 011407.

\bibitem{DLA-MSV} A. Yu. Menshutin, L.N. Shchur, V.M. Vinokur, Phys. Rev. E {\bf 75} (2007) 010401(R).

\bibitem{BB-alg} R.C. Ball and R.M. Brady, J. Phys. A {\bf 18} (1985)
L809.

\bibitem{TM-alg} S. Tolman and P. Meakin, Physica A {\bf 158} (1989)
801; Phys. Rev. A {\bf 40} (1989) 428.

\bibitem{MBRS-alg} P. Meakin, R.C. Ball, P. Ramanlal, L.M. Sander
{\bf 35} (1987) 5233.

\bibitem{SSZ} E. Sander, L. M. Sander, R. M. Ziff, Computers in Physics {\bf 8}
(1994), 420-425; L.M. Sander, Contemporary Physics {\bf 41} (2000),
203-218.

\bibitem{KVMW} H. Kaufman, A. Vespignani, B.B. Mandelbrot, and L.
Woog, Phys. Rev. E {\bf 52} (1995) 5602.

\bibitem{book} S. Redner, {\it Guide to First-Passage Processes}
 (Cambridge University Press, New York, 2001).

\bibitem{Mandelbook} B.B. Mandelbrot, {\it The fractal geometry of nature} (W.H. Freeman and Company, New York, 1982)

\bibitem{ms-PZ} M. Plischke and Z. R\'acz, Phys. Rev. Lett. {\bf 53}
(1984)  415.

\bibitem{nms-SBBS} E. Somfai, R.C. Ball, N.E. Bowler, L.M. Sander,
Physica A {\bf 325} (2003) 19.

\bibitem{Lee} J. Lee, S. Schwarzer, A. Coniglio, H.E. Stanley,
Phys. Rev. E {\bf 48} (1993) 1305.

\bibitem{Hastings97} M.B. Hastings, Phys. Rev. E {\bf 55} (1997) 135.

\end{thebibliography}
\end{document}